\documentstyle[11pt,aas2pp4,flushrt,psfig]{article}

\def\he3{\hbox{He $3p$}}
\newcommand\etal{et~al.}
\newcommand\xray{X-ray}

\newcommand\psrn{PSR~J1105$-$6107}
\newcommand\msh{MSH 11$-$61A}
\newcommand\asca{{\it ASCA}}
\newcommand\egret{{\it EGRET}}
\newcommand\rosat{{\it ROSAT}}

\newcommand\gro{{\it CGRO}}

\newcommand{\goe}{\stackrel{>}{\sim}}

\slugcomment{To appear in the Astrophysical Journal Letters}

\begin{document}

\title{\Large\bf X-Ray Emission from the Radio Pulsar \psrn}

\author{E. V. Gotthelf\altaffilmark{1}}
\affil{Laboratory for High Energy Astrophysics, NASA/GSFC, Greenbelt, MD 20771}
\authoremail{gotthelf@gsfc.nasa.gov}

\author{V. M. Kaspi}
\affil{Department of Physics and Center for Space Research, Massachusetts Institute of Technology, Cambridge, MA, 02139}
\authoremail{vicky@space.mit.edu}

\altaffiltext{1}{\rm Universities Space Research Association}

\begin{abstract}

We have detected significant \xray\ emission from the direction of the
young radio pulsar \psrn\ using the \asca\ Observatory.  The $11
\sigma$ detection includes $460$ background-subtracted source counts
derived using data from all four \asca\ detectors. The emission shows no
evidence of pulsations; the pulsed fraction is less than 31\%, at the 90\%
confidence level. The \xray\ emission can be characterized by a
power-law spectrum with photon index $\alpha = 1.8 \pm 0.4$, for a
neutral hydrogen column density N$_H = 7 \times 10^{21}$~cm$^{-2}$. The
unabsorbed $2-10$ keV flux assuming the power-law model is ($6.4 \pm 0.8) 
\times 10^{-13}$ ergs cm$^{-2}$ s$^{-1}$. The implied efficiency
for conversion of spin-down luminosity to \asca-band emission is
($1.6 \pm 0.2)\times 10^{-3}$, assuming a distance of 7~kpc to the
source.  Within the limited statistics, the source is consistent with
being unresolved.  We argue that the X-rays are best explained as
originating from a pulsar-powered synchrotron nebula.

\end{abstract}
\keywords{stars: neutron --- pulsars: individual: (PSR J1105$-$6107) ---
supernova remnants --- X-rays: stars}

\section{Introduction}

X-ray observations of radio pulsars provide a powerful probe of the
energetics and emission mechanisms of rotation-powered neutron stars.
Though pulsars have traditionally been most easily studied at radio
wavelengths, that emission represents a small fraction of the neutron
star energy budget, which is fed by rotational energy loss due to
magnetic dipole braking.  It is thought that the bulk of this
``spin-down luminosity'' emerges in the form of a relativistic wind.
The interaction of the wind with the surrounding medium can lead to
observable synchrotron emission in the X-ray band, as in the Crab
Nebula.  Thermal X-rays are also detected from the neutron star surface
in some cases, as are pulsed, non-thermal X-rays from the
magnetosphere.  Recently the number of X-ray identified
rotation-powered pulsars has grown sufficiently large that studying the
properties of the population as a whole is possible (see Becker \&
Tr\"{u}mper 1997 for a recent review).  Particularly important targets
for X-ray observations are the youngest members of the pulsar
population, because of their large spin-down luminosities.

The recently discovered young radio pulsar \psrn\ has a spin period
63~ms, a characteristic spin-down age $\tau = 63$~kyr, and a spin-down
luminosity $\dot{E}=2.5\times$10$^{36}$~erg~s$^{-1}$ (Kaspi \etal\
1997).  The pulsar's dispersion measure implies that it is 7~kpc from
the Earth (Taylor \& Cordes 1993).  \psrn\ is positionally coincident
with the high-energy $\gamma$-ray \gro/\egret\ source
2EG~J1103$-$6106. If the two are associated, the observed $\gamma$-ray
flux implies an efficiency
for conversion of spin-down luminosity to $\gamma$-rays of $\sim$3\%
for a beaming angle of 1~sr.  The pulsar is
located $\sim$25$^{\prime}$ from the center of the supernova remnant
\msh, which suggests that the two may be associated.  At 7~kpc,
if the characteristic age is the true age (but see Kaspi \etal\ 1997
for a discussion) and if the apparent geometric center of the remnant
is the approximate pulsar birth location, the pulsar's implied
transverse velocity is $\sim$650~km~s$^{-1}$.  This velocity is high,
but within the range of typical young pulsar velocities (Lyne \&
Lorimer 1994).  Nevertheless, the large uncertainties in the age and
distance estimates toward both sources and absence of independent
evidence for a large velocity make an association tentative
at best (c.f. Kaspi 1996).

We report here on \asca\ observations of the \psrn\ field, which 
reveal a faint X-ray source in the direction of the pulsar. 

\section{Observations}

A two-day-long observation of \psrn\ was performed by the \asca\
Observatory (Tanaka \etal\ 1994) on 1996 June 21--23. Data were acquired
with all four imaging spectrometers, each in the focal plane of its
own foil mirror: two Solid State Imaging Spectrometers (SIS-0, SIS-1),
and two Gas Imaging Spectrometers (GIS-2, GIS-3). These spectrometers
offer moderate energy ($\sim 5\%$) and $\sim 2^{\prime}$ imaging
resolution in their $ \sim 1.0-10$ keV energy band-passes.  The GIS
has a higher effective area above $\sim 2$ keV and a greater net
observation time than the SIS.

To search for pulsations from \psrn, GIS data were collected in the
highest time resolution configuration ($0.488$, $3.906$, or $15.625 \ \rm{ms}$ depending on data acquisition rate). SIS data were acquired
in 2 CCD mode, with 8 s integrations, using a combination of FAINT and
BRIGHT modes (see the \asca\ Data Analysis Guide for details). The
target was placed $3^{\prime}$ off the corner of SIS-0 CCD-0, as close
to the mean telescope optical axis as was practical, to minimize
vignetting losses. The data were edited to exclude times of high
background contamination using the standard ``REV2'' screening
criteria. This rejects time intervals of South Atlantic Anomaly
passages, Earth block,
bright Earth limb in the field-of-view, and periods of high particle
activity. The resulting effective observation time is $2 \times
56$~ks (GIS) and $2 \times 38$~ks (SIS) for the two pairs of
sensors. The CLEANSIS algorithm (Gotthelf 1993) was used to identify
and remove hot and flickering CCD pixels in the SIS data.

\section{Image}

Flat-fielded images were generated by aligning and co-adding
exposure-corrected images from the pairs of instruments.  Exposure
maps were generated with ASCAEXPO, \asca\ software which uses the
satellite aspect solution, instrument map (GIS), chip alignment, and
hot pixel map (SIS) to determine the exposure time for each sky image
pixel. The exposure correction was highly effective in removing GIS
instrumental structure due to the window support grid. Figure 1a
displays the resultant smoothed broad-band image for the GIS. The
image reveals the supernova remnant towards the northern edge of the
field-of-view, an unresolved, serendipitous point source close to the
Eastern edge, and a faint source near the pulsar position.

We performed an unbiased source search on the GIS data using a
modified box search algorithm set for a weak source threshold.  For
each test cell on the sky image we fit the putative source profile to
a model for the PSF plus background; the exposure map is used to
compute the statistical errors of the fit.  We detected both
the faint source and the serendipitous object. The only other sources
detected by this search were the significant flux deviations associated
with parts of the remnant in the north. Here we discuss the putative
target detection and leave the analysis of the serendipitous source,
\hbox{AXS~1106$-$6103}, to a future paper.

The measured \xray\ position of the faint source is $0.8^{\prime}$
from the radio position of \psrn\ (see Table 1). This is consistent
with the \asca\ error circle (Gotthelf 1996), which introduces a
systematic offset to the image in a random sky direction. This offset
dominates over the source position measurement error of $\sim
15^{\prime\prime}$. We estimate that the probability of detecting a
source with comparable flux by chance in the GIS active area, which
has a radius of $22^{\prime}$, to be $< 0.008$.  For this estimate we 
have assumed there is $\sim1$ such source per deg$^2$ and that the spatial
resolution is $2^{\prime}$. We therefore consider this a detection of
\psrn\ and proceed to estimate the detection significance and source
flux.

To estimate the significance of the detection, we ignore the
complexities of the \asca\ point-spread function and compare the
number of
counts collected from an optimal 4$^{\prime}$ diameter aperture centered on the
source with that from a $5^{\prime} - 11^{\prime}$ concentric
annulus.  The relatively small number of counts available in the
source region makes the approximation reasonable. The total number of
counts in the GIS source aperture is 840, of which 616 are expected
from the background, implying a $7.5\sigma$ pulsar detection. We have
normalized the counts in the background annulus (13,400 cts) to the
source aperture by the relative exposure area ($\beta = 0.046$).
The source is also detected in the individual GIS sensors, but at the 
expected lower significance (see Table 2).  Our estimates of the detection
significance (the signal-to-noise ratio, $S/N$) 
take into account both the source and background variance:
\begin{equation}
 S/N = S / \sqrt{S + B ( 1 + 2 \beta) }, 
\end{equation}
where $S = N_s - B$ is the number of background subtracted source
counts, derived from $N_s$, the total number of counts in the
aperture, and $B = N_b \beta$, the number of background counts,
normalized to the source aperture by the relative exposure area $\beta
= T_s / T_b$.  Here, $T_{s,b} = \Sigma_{s,b} \ T_{i,j}^{expo}$ gives
the total exposure time by summing the exposure map ($i,j$) pixels
covering the source or background aperture. Notice that if all pixels
contain the same exposure time, then $\beta$ is equivalent to the
usual area ratio of the source to background aperture.

We next examined the flat-fielded SIS image, similarly co-added and
smoothed on arcminute scales as with the GIS. There is also evidence
of enhanced emission around the location of the pulsar, within the
expected error circle (see Fig. 1b). The detected number of counts at the
putative pulsar position is too small to fully resolve the familiar
cross pattern of the PSF, as the morphology is dominated by Poisson
fluctuations (see Hwang \& Gotthelf 1997 \S 2.2 for a discussion of the
significance of peaks in similarly processed images). It is reassuring
that the brightest pixel in the SIS image is coincident with the GIS
location of \psrn. Using the same method outlined for the GIS we
derive a $7.7 \sigma$ SIS detection from 235 background subtracted
source counts.

Thus, including data from all four telescopes aboard \asca, we find
a total of 459 background-subtracted source counts from the direction
of \psrn, with a significance of detection of 10.8$\sigma$.

We note that there is ambiguous evidence for extended emission between the
pulsar and remnant.  Also, there is perhaps another faint point source within a
few arcminutes of the pulsar. This additional flux is evidently very soft
as it is not seen in the GIS image and is confined to the soft-band SIS
image ($< 2$ keV).  However, it is not evident in archive \rosat\
PSPC data of the region. Another faint point source, below the GIS
broad-band source detection threshold, is apparent in the soft-band
SIS image just south of the pulsar and does correspond to faint PSPC 
emission.

\subsection{\bf Spectrum}

Although the number of source counts from the pulsar is too small to
distinguish among spectral models, we assume a simple absorbed
power-law model to characterize the spectrum and extract a flux
measurement.  We concentrate our spectral analysis on the GIS, rather
than SIS, as the former has
greater sensitivity to photons above $2$ keV and has a
greater field-of-view for background estimation.  Source counts from GIS-2
and GIS-3 were extracted using a $6^{\prime}$ diameter circular region
and the fits were restricted to the $1 - 10$ keV energy
range. Background counts were obtained from a region $14^{\prime}$
west of the source region. This is far enough to provide a background
dominated by local emission but still within the region of enhanced
emission, most likely similar to the source background.

We estimate the equivalent neutral hydrogen column density toward the
X-ray source using the measured value toward \msh, N$_H \simeq (7 \pm
2) \times 10^{21}$~cm$^{-2}$ (Rho 1995).  This is consistent with the
value estimated using the standard ``rule-of-thumb'' of 10 neutral
hydrogen atoms per free electron (Seward \& Wang 1988), $7 \times
10^{21}$~cm$^{-2}$, given the pulsar dispersion measure, $271.01 \pm
0.2$~pc~cm$^{-3}$ (Kaspi \etal\ 1997).  Radio observations of the
field show that there is no large line-of-sight HII region that could
contribute significantly to the dispersion measure (B. Stappers,
B. Gaensler, personal communication).  A fit to the
background-subtracted spectrum with a power-law model using the above
column density yields a photon index of $\alpha = 1.8 \pm 0.4$.  The
implied unabsorbed $2 - 10$ keV flux is ($6.4 \pm 0.8) \times 10^{-13}$
ergs cm$^{-2}$ s$^{-1}$.  The predicted SIS count rates for the above
model are consistent with the SIS detection.

As discussed below, the pulsar's X-ray luminosity is important for
interpreting the origin of the X-ray emission.  Therefore, we test the
robustness of our flux measurement in face of the uncertain column
density.  Allowing both the power-law index and
column density to vary, we find
best fit parameters of $\alpha = 1.7 \pm 0.3$ and N$_H \sim
0.6_{0.0}^{1.8} \times 10^{22}$ cm$^{-2}$. The implied $2 - 10$ keV
flux of $(6.5 \pm 0.8) \times 10^{-13}$ ergs cm$^{-2}$ s$^{-1}$ is
statistically unchanged from the previous fit.  We also held N$_H$
fixed at half and twice the nominal value of $7 \times
10^{21}$~cm$^{-2}$.  With half the hydrogen column, neither the implied
flux ($6.6 \pm 0.8 \times 10^{-13}$ ergs cm$^{-2}$ s$^{-1}$) nor the
power law index ($\alpha = 1.5 \pm 0.3$) differs significantly.
With twice the column density, the best-fit power law index is $\alpha
= 2.2 \pm 0.4$, implying a flux of $(6.4 \pm 0.8) \times 10^{-13}$ ergs
cm$^{-2}$ s$^{-1}$, again not significantly different.  We conclude
that our flux measurement is insensitive above $>2$ keV to the assumed
N$_H$, within the reasonable range explored. 

\subsection{\bf Timing}

We carried out a timing analysis using the events recorded by the two
GIS instruments.  We selected events from a 4$^{\prime}$ diameter
aperture centered on the source, using data acquired at the high and
medium data rates only. This produced total of $763$ events.  We then
folded the events using an ephemeris obtained from radio timing
observations of \psrn\ at the Parkes radio telescope. The ephemeris,
valid from 1996~May~10 through 1996~July~4, consists of the following
parameters: $f = 15.8247373445677$~s$^{-1}$, $\dot{f}= -3.966465
\times 10^{-12}$~s$^{-2}$, $\ddot{f}= -5.32 \times 10^{-25}$~s$^{-3}$,
referred to epoch MJD~50240.0, where $f$ is the rotation frequency.
No pulsation was apparent in the folded
light curve.  To set an upper limit to the pulsed fraction, we
injected an artificial pulsed signal of variable pulsed fraction and
duty cycle 0.5, until it was detected at the 90\% confidence level
using a standard $\chi^2$ test.  Our upper limit to the pulsed
fraction determined in this way is 0.31.

\section{\bf Discussion}

Because \psrn\ has a characteristic age $\tau= 63$~kyr, thermal
emission from the surface should have effective temperature no higher
than 100~eV (see \"{O}gelman 1995 and references therein). The
corresponding blackbody flux, assuming the 7~kpc distance and a
neutron star radius of 10~km, is 2.3$\times
10^{-13}$~erg~cm$^{-2}$~s$^{-1}$.  For this spectrum and flux, the
\asca\ GIS count rate should be no more than $\sim$4$\times
10^{-5}$~cps for reasonable values of $N_H$.  This count rate is some
two orders of magnitude below the observed rate (see Table~2).  Thus,
thermal emission cannot be the origin of the X-rays we detect from
\psrn.

The most energetic pulsars in the known population exhibit
hard-spectrum pulsations due to emission from the magnetosphere.  Such
pulsations are characterized by large pulsed fractions (Becker \&
Tr\"{u}mper 1997 and references therein), much larger than the upper
limits we set using our \asca/GIS data for \psrn.  This is not
surprising given the characteristic age of \psrn, which suggests it
will exhibit properties like those of Vela-type pulsars, which do not
show strong X-ray pulsations (e.g. \"{O}gelman, Finley \& Zimmerman
1993).  Deeper observations could reveal pulsations due to a small
magnetospheric component, but the bulk of the observed flux, being
unpulsed, must have a separate origin.

The observed X-ray emission could originate from an optically thin hot
gas bubble emitting thermal bremsstrahlung radiation, possibly a
faint, previously unknown supernova remnant.  If $\tau= 63$~kyr is
the system's true age, the hypothetical remnant would probably be in
the radiative phase.  Then, the remnant radius, $R \simeq 28.5
(E_{51})^{5/21} n_0^{-5/21} t_5^{2/7} \;\; {\rm pc},$ where $E_{51}$
is the supernova initial energy in units of $10^{51}$~erg, $n_0$ is
the initial surrounding interstellar medium number density, and $t_5$
is the age of the remnant in units of $10^5$~yr (see Lozinskaya 1992
and references therein).  For $t_5 = 0.63$ and taking $R < 6$~pc,
corresponding to the maximum size of the remnant given the \asca/GIS
spatial resolution and assuming a distance of 7~kpc, $E_{51}/n_0 < 3
\times 10^{-3}$.  Thus either this supernova released far less energy
than is typical, or it expanded into a region of density
$\sim$300~cm$^{-3}$.  Neither seems likely.  Furthermore, there is no
evidence for radio emission from a putative remnant (B. Stappers, B. Gaensler,
personal communication).  It is possible that the observed X-ray
emission comes from only a small part of a larger region, whose
surface brightness is too faint to be observed.  Only a spectrum and
deeper observations can decide unambiguously.

The likeliest origin for the X-rays detected from \psrn\ is a
synchrotron nebula powered by the pulsar via a relativistic wind.  The
most famous pulsar wind nebula surrounds the Crab pulsar (Rees \& Gunn
1974; Kennel \& Coroniti 1984; Emmering \& Chevalier 1987; Gallant \&
Arons 1994).  There the pulsar wind pressure is presumably confined 
by the unseen expanding shell.  In pulsar wind nebulae, the
wind relativistic electrons and positrons (and possibly heavy ions,
Hoshino et al. 1992) are confined, accelerated at the reverse shock,
and radiate synchrotron emission.
For PSR~J1105$-$6107, given the absence of evidence for a confining
shell, the pulsar wind is probably confined
by ram pressure (Cheng 1983; Kulkarni et al. 1992; Fruchter et al. 1992; Wang,
Li, \& Begelman 1993; Finley, Srinivasan, \& Park 1996; Romani, Cordes
\& Yadigaroglu 1997).

Here we show that a ram-pressure confined synchrotron nebula is consistent
with the \asca\ detection of PSR~J1105$-$6107.
Equating the interstellar medium ram pressure with that of the pulsar
wind, assuming a thin shock for simplicity, the distance of the shock apex from the pulsar 
is given by
\begin{equation}
r_s = \left( \frac{f\dot{E}}{4 \pi c \rho v^2} \right)^{1/2},
\end{equation}
where $f$ is the fraction of $\dot{E}$
channeled into the wind, $\rho$ is the ambient
density, and $v$ is the pulsar velocity.  It is reasonable to assume
$f \simeq 1$ given the low efficiency for conversion to high-energy
$\gamma$-rays (Kaspi et al. 1997).  Let $\rho_1 \equiv 1$~H
atom~cm$^{-3}$ and $v_{100} \equiv (v/100$~km~s$^{-1}$).  Then $r_s
\simeq 2 \times 10^{17} \rho_1^{-1/2} v_{100}^{-1}$~cm.  The magnetic
field just upstream of the shock, $B_u$, is given by
\begin{equation}
B_u = \left( \frac{\sigma f \dot{E}}{(1+\sigma)r_s^2 c}\right)^{1/2}
\end{equation}
where $\sigma$ is the ratio of the magnetic energy flux to the kinetic
energy flux of the pulsar wind upstream of the shock.  Observations of
the Crab pulsar show that $\sigma \simeq 0.005$ (Kennel \& Coroniti
1984).  Just downstream of the shock, $B_d = 3B_u$ from shock jump
conditions (e.g. Arons \& Tavani 1993).  We find $B_d \simeq 1 \times
10^{-5} \sigma_{0.005}^{1/2} \rho_1^{1/2} v_{100}$~G, where
$\sigma_{0.005} \equiv \sigma/0.005$.  The pulsar wind flow time scale
is $t_f \sim 3r_s/c \simeq 2 \times 10^7 \rho_1^{-1/2}
v_{100}^{-1}$~s.  The time scale for synchrotron cooling is given by
$t_s(\gamma) = 3 m^3 c^5/2e^4 B_d^2
\gamma = 5.5 \times 10^{18} \sigma_{0.005}^{-1} \rho_1^{-1}
v_{100}^{-2} \gamma^{-1}$~s, where $\gamma$ is the post-shock pair
Lorentz factor (e.g. Rybicki \& Lightman 1979).
The synchrotron cooling efficiency for pairs having
$\gamma \sim 10^8$ is therefore $\epsilon_8 \equiv t_f/t_s(\gamma_8) =
3.6 \times 10^{-4} \sigma_{0.005} \rho_1^{1/2} v_{100} \gamma_8$,
where $\gamma_8 \equiv \gamma/10^8$.  The bolometric luminosity in
synchrotron emission $L_s$, which is dominated by the highest energy
pairs, is given for $\gamma_8$ pairs by $L_s = \epsilon_8 f_8
\dot{E}$, where $f_8$ is the fraction of pairs having $\gamma = 10^8$.
This predicted bolometric luminosity is 4--5 times
smaller than the observed \asca-band luminosity alone.  
Relaxing the assumption that $f = 1$ exacerbates the problem.
However, the observations can be accounted for in
a number of ways:  (i) if all of the
post-shock wind pairs have Lorentz factors $\gamma >> 10^8$ or $\sigma$ is
significantly larger than 0.005, implying
different wind properties from those inferred from the Crab
pulsar (e.g. Hoshino et al. 1992; Gallant \& Arons 1994);
(ii) the thin shock assumption does not apply and we have underestimated
$t_f$;
(iii) the pulsar resides in an overdense region having
$\rho_1 \goe 20$~cm$^{-3}$;
(iv) the actual space velocity is larger than assumed above, i.e. $v_{100} \goe 5$.
Case (iv) would be consistent with estimates of the pulsar velocity distribution
(e.g. Lyne \& Lorimer 1994), and with the velocity implied on the basis of a possible
association with the supernova remnant \msh\ (Kaspi et
al. 1997).   Estimates of the SNR shock velocity (P. Slane, personal
communication) suggest an even higher pulsar space velocity (c.f. Frail
et al. 1996).  Extrapolating the \asca-band flux to the
\rosat\ band assuming a power-law index of $-2$, we find a soft X-ray
luminosity of PSR~J1105$-$6107 larger by nearly an order of magnitude compared with 
those of other pulsars of comparable age and $\dot{E}$.  If similar pulsars have
similar wind properties, this suggests differing environments and/or space
velocities.  Caution is of course
required given the systematic uncertainties due to the distance model
and poorly determined X-ray spectra.

Improved spectral measurements and high spatial resolution X-ray
imaging of the source may distinguish among the above possibilities.
For the faint thermal remnant model or cases (i) and (ii) above,
the emission should be
spatially resolvable by {\it AXAF}.  For a high velocity pulsar, the
shock apex distance would not be resolvable, but the swept back
extended bow shock ``tail'' morphology might be apparent.

\bigskip

\acknowledgments 

We thank M. Bailes, R. Manchester and R. Pace for acquiring radio timing data
for \linebreak \psrn, and B. Gaensler, P. Slane, and B. Stappers for discussing 
data prior to publication.  We thank J. Arons, M. van Putten, G. Vasisht, and Q. D. Wang 
for useful discussions.  This work was supported by NASA grant NAG-51436.  V.M.K.
received additional support from Hubble Fellowship grant number
HF-1061.01-94A from the STScI, which is operated by the AURA, under
NASA contract NAS5-26555. E.V.G. is supported by USRA under NASA
contract NAS5-32490.

\onecolumn

\begin{deluxetable}{l c c}
\small
\tablewidth{0pt}
\tablecaption{J2000 Source Positions}
\tablehead{
 \colhead{Source} & \colhead{Right Ascension}  & \colhead{Declination}
}
\startdata
    \psrn\ (radio)\tablenotemark{a}    & $11^h 05^m 26.07^s$ & $-61^{\circ} 07^{\prime} 52.1^{\prime\prime} $ \nl
    \psrn\ (\asca)                     & $11^h 05^m 19^s$ & $-61^{\circ} 07^{\prime} 55^{\prime\prime} $\nl
\enddata
\tablenotetext{a}{The radio position is uncertainty by 0.7$^{\prime\prime}$.}
\end{deluxetable}

\begin{deluxetable}{c c c c c c}
\small
\tablewidth{0pt}
\tablecaption{\asca\ detection of \psrn}
\tablehead{
\colhead{Sensor} & \colhead{Exposure} & \colhead{Count Rate\tablenotemark{a}}   & {Bgd Rate\tablenotemark{b}}     & \colhead{Significance\tablenotemark{c}} \\
              & \colhead{(ksec)}      & \colhead{($ \times 10^{-3}$ cps)}       & \colhead{($ \times 10^{-3}$ cps)}  & \colhead{($\sigma$)} 
}
\startdata
  SIS-0 & 38.9 & 12.9 &  9.5 &  5.8 \nl
  SIS-1 & 37.8 & 10.5 &  7.9 &  4.9 \nl
  GIS-2 & 55.9 &  6.6 &  5.2 &  4.1 \nl
  GIS-3 & 55.8 &  8.3 &  5.8 &  6.5 \nl
  Total &188.4 &  --  &  --  & 10.8 \nl
\enddata
\tablenotetext{a}{Total source plus background count rate in a $4^{\prime}$ diameter aperture centered on the \asca\ PSR~J1105$-$6107 position.}
\tablenotetext{b}{Background count rate in a $5^{\prime}-11^{\prime}$ annulus concentric with the source position and normalized to the source aperture (see text). The background annulus for the SIS is truncated by the CCD boundary.}
\tablenotetext{c}{See text for definition.}
\end{deluxetable}

\clearpage

\begin{figure}

\centerline{
\psfig{figure=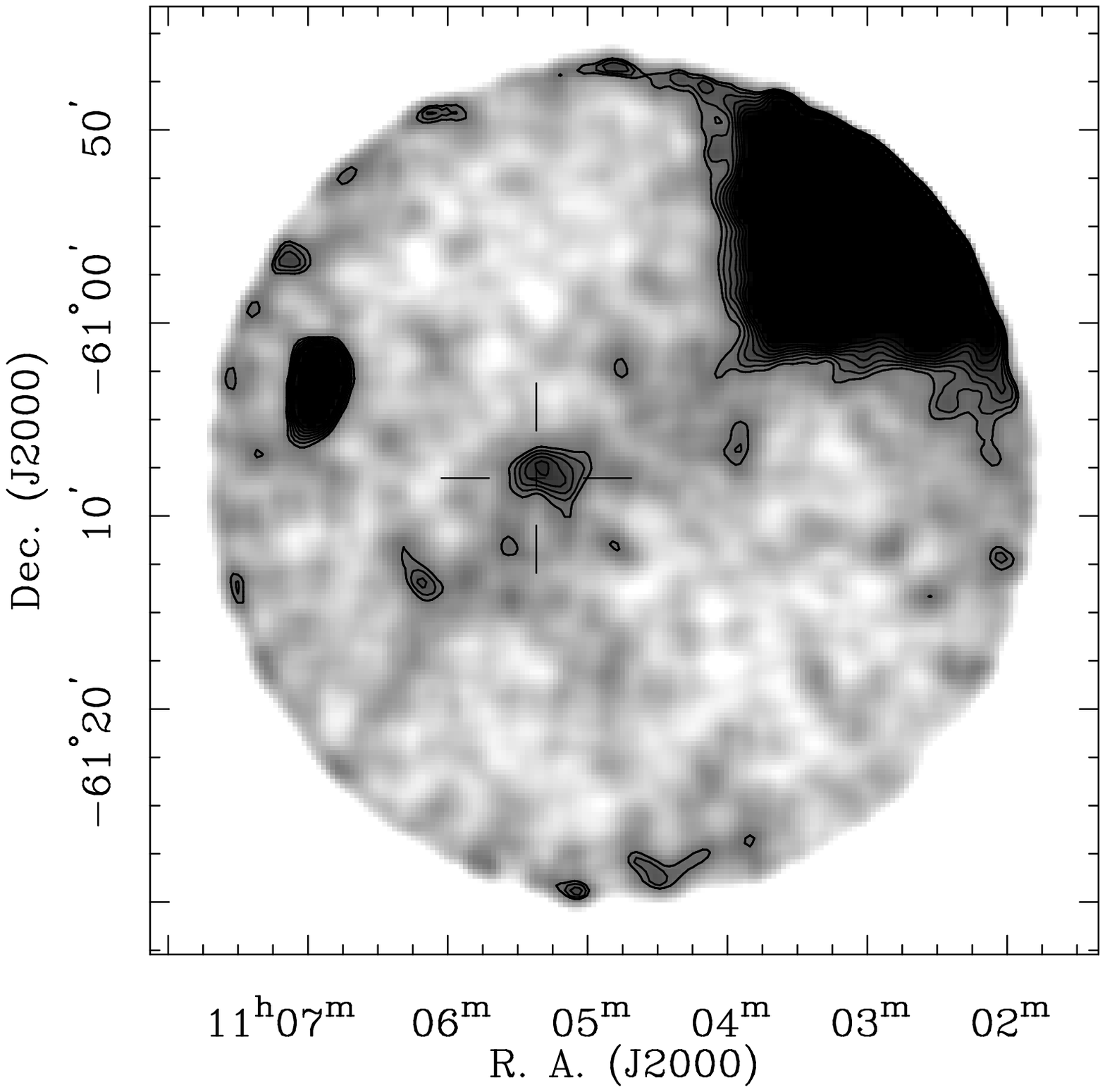,height=4truein,angle=270.0,bbllx=25bp,bblly=25bp,bburx=587bp,bbury=500bp,clip=}
\psfig{figure=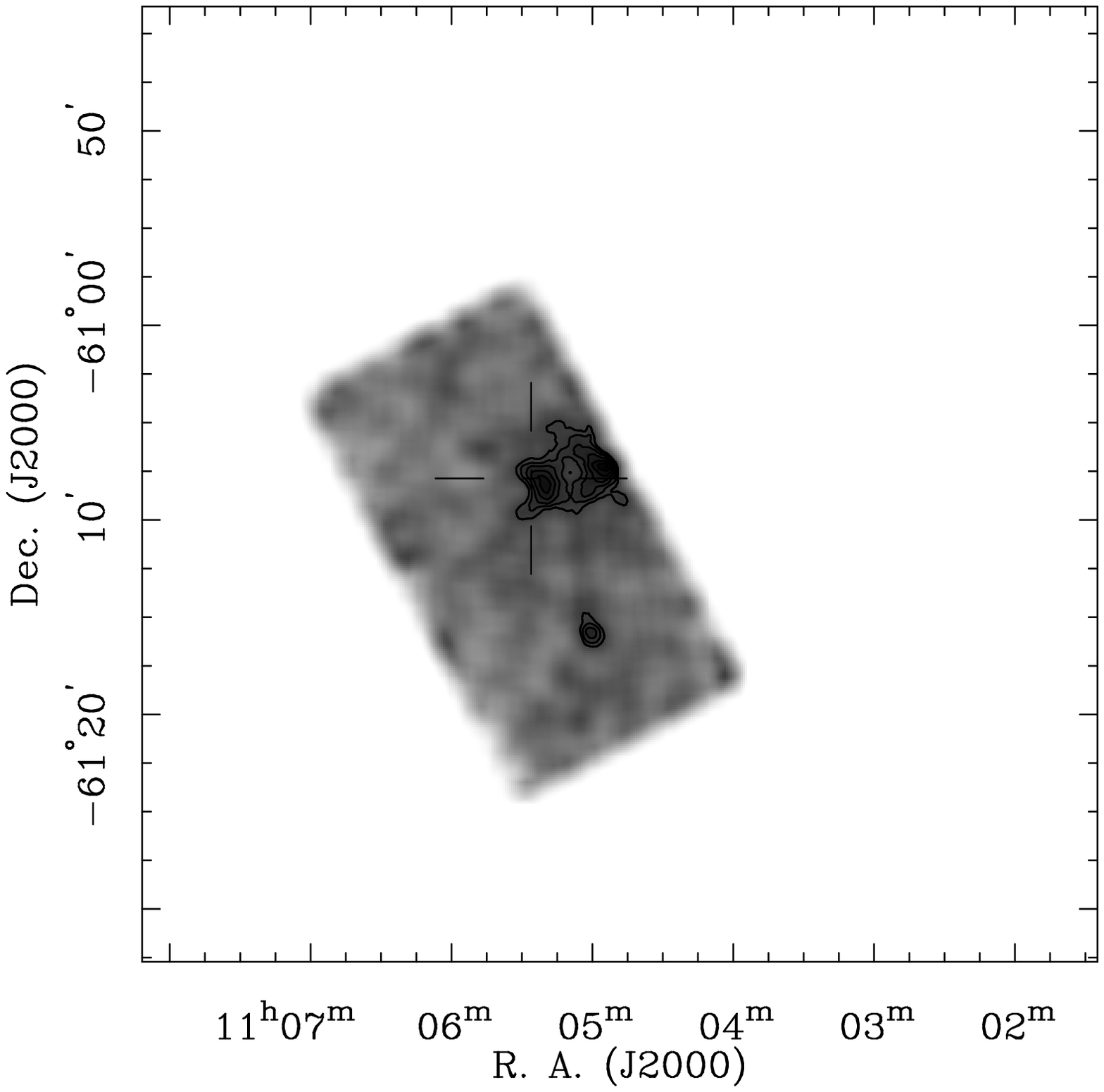,height=4truein,angle=270.0,bbllx=25bp,bblly=25bp,bburx=587bp,bbury=500bp,clip=}
}
\caption{\asca\ images of the \psrn\ field: flat-fielded images
of the region around the pulsar, whose location is marked by the
cross.  (Left) The broad-band GIS image shows X-ray emission coincident with the pulsar,
supernova remnant \msh\ to the north, and an unknown serendipitous source to the west. 
The image is saturated to highlight the pulsar emission. (Right) The broad-band 
SIS image plotted on the same scale as the GIS. The two additional sources of significant
flux are confined to the soft energy band. The contours are at the $3\sigma, 4\sigma, 
5\sigma, 6\sigma, 7\sigma\ {\rm and} \; 8\sigma$ levels; the GIS image is clipped at the 
$12 \sigma$ level for clarity.} 
\end{figure}

\end{document}